\documentstyle[12pt]{article}

\newcommand{\rf}[1]{(\ref{#1})}
\newcommand{\bea}{\begin{eqnarray}}
\newcommand{\eea}{\end{eqnarray}}
\newcommand{\beas}{\begin{eqnarray*}}
\newcommand{\eeas}{\end{eqnarray*}}
\newcommand{\nn}{\nonumber\\ }
\newcommand{\g}{\gamma}
\renewcommand{\l}{\lambda}
\renewcommand{\b}{\beta}
\renewcommand{\a}{\alpha}
\newcommand{\n}{\nu}
\newcommand{\m}{\mu}

\newcommand{\ep}{\varepsilon}

\newcommand{\dl}{\delta}
\newcommand{\prt}{\partial}

\newcommand{\oh}{\frac{1}{2}}

\newcommand{\cD}{{\cal D}}

\newcommand{\ra}{\right\rangle}
\newcommand{\la}{\left\langle}
\newcommand{\ha}{\hat{a}}
\newcommand{\hg}{\hat{g}}

\def\void{}
\def\labelmark{}

\newenvironment{formula}[1]{\def\labelname{#1}
\ifx\void\labelname\def\junk{\begin{displaymath}}
\else\def\junk{\begin{equation}\label{\labelname}}\fi\junk}%
{\ifx\void\labelname\def\junk{\end{displaymath}}
\else\def\junk{\end{equation}}\fi\junk\labelmark\def\labelname{}}

\newcommand{\beq}{\begin{formula}}
\newcommand{\eeq}{\end{formula}}
\newcommand{\beqv}{\begin{formula}{}}

\begin{document}
\topmargin 0pt
\oddsidemargin 5mm
\headheight 0pt
\topskip 0mm

\hfill NBI-HE-93-63

\hfill November  1993

\hfill hep-th/9312002

\addtolength{\baselineskip}{0.20\baselineskip}
\begin{center}

\vspace{36pt}
{\large \bf 2d quantum gravity coupled to \\  renormalizable matter fields}

\end{center}

\vspace{36pt}

\begin{center}
{\sl Jan Ambj\o rn} and {\sl Kazuo Ghoroku}\footnote{Permanent address:
Department of Physics, Fukuoka Institute of Technology,
Wajiro, Higashi-ku, Fukuoka 811-02, Japan}

\vspace{12pt}

The Niels Bohr Institute,\\
 Blegdamsvej 17, DK-2100 Copenhagen \O , Denmark

\end{center}

\vspace{36pt}

\vfill

\begin{center}
{\bf Abstract}
\end{center}

\vspace{12pt}

\noindent
We consider two-dimensional quantum gravity coupled to matter fields
which are renormalizable, but not conformal invariant. Questions
concerning the $\b$ function and the effective action are addressed,
and the effective action and the dressed renormalization group equations
are determined for various matter potentials.

\vspace{24pt}

\vfill

\newpage

\section{Introduction}

In dimensions higher than two the quantum theory of gravity
poses a major problem. The problem of gravitational dressing of renormalizable
field theories seems minor in comparison. In two dimensions we understand
by now how to quantize gravity in the absence of matter,
and we understand the gravitational
dressing of conformal field theories (at least if $c \leq 1$). However,
the gravitational dressing of  a general renormalizable two-dimensional
field theory is still not well understood. This problem has become
increasingly important in the study of two-dimensional black holes,
which again serve as interesting toy models for higher dimensional theories.

In the next section we briefly summarize the present knowledge of
how to  perturb away from a conformal field theory coupled
to matter. If we work in conformal gauge the quantized theory of gravity
and matter must be conformal invariant with respect to the fiducial
metric introduced in the decomposition $g_{\a\b} \to \hg_{\a\b} e^{2\rho}$.
This is the statement that the $\b$-function must be zero when the
free energy is viewed as a function of the coupling constants and the
cut-off associated with the fiducial metric. The requirement of a
vanishing $\b$-function allows us to determine the back-reaction of
quantum gravity on the matter fields to  second order
in the  coupling constants when
we perturb the matter theory away from a conformal fixed point by
marginal perturbations.

We can view the combined theory of Liouville
and matter as a generalized non-linear sigma model with the constraint
imposed that the $\b$-function has to vanish. As is well known
from the usual critical string theory the vanishing of the $\b$-function
of this generalized sigma model can be formulated as a set
of equations in target space. These can be viewed as the classical
equations of motion for a certain action in target space.
Formulated this way our task is to solve the classical equations,
and the solution will determine the interaction between Liouville and
matter fields. This approach is in certain respects to be preferred to
more direct approach outlined in sec. 2 since it is better suited
for making a non-perturbative ansatz. It is however plagued by
ambiguities due to the choice of boundary conditions for the partial
differential equations. These ambiguities reflect that infinite many
counter terms can be added to the non-linear sigma model.
Several attempts have made to restrict the allowed boundary conditions
of the classical equations \cite{3}. In this article we propose
to solve the equations perturbatively with respect to the coupling
constants which takes the matter theory away from a conformal field
theory.  Since we know the solution at the conformal point we will
have no ambiguity in the solutions of the classical equations if we
impose that they should go smoothly over into the corresponding solutions
at the conformal point. In this way we can calculate the effective
action in  concrete models and verify the ``gravitational dressing''
of the $\b$-function mentioned above.

The rest of the paper is organized as follows: In sect. 2 we discuss the
gravitational back-reaction on matter when we move away from
a conformal matter theory. The universal change of the $\b$-function
is found. In sect. 3 we review how to reformulate the problem as
classical equations of motion of a certain action in target space,
and in sections 4 to 6  we solve perturbatively these equations in
specific models. Finally section 7 contains a discussion of the results.

\section{Matter coupled to gravity}

\subsection{First order corrections to DDK}

We know at the formal level how to couple conformal invariant
field theories to 2d quantum gravity. The results of KPZ \cite{kpz}
and DDK \cite{ddk} solves the problem for $c<1$ and (at a formal level)
for $c>25$, where $c$ denotes the conformal charge of the matter
theory before coupling to gravity. We even know the correct way to
discretize the coupled model \cite{david1,adf,kkm} and to
look for critical points in the corresponding statistical system,
thereby reaching contact with the continuum results of KPZ and DDK.
However, in a statistical mechanics context the critical point and the
corresponding conformal field theory are singular, albeit interesting
points. The interest is usually centered around the approach to the
critical point, as  governed by the renormalization group. This
approach allows us to identify masses and coupling constants of the
class of theories associated with the critical point. The concept
of distance plays an important role when discussing the renormalization
group and masses. In a theory where we integrate over all metrics the
most naive definition of correlation functions involves the  use of the
geodesic distance $r$ for each metric $g$, the calculation of  the correlation
function $\Delta_g (r)$ (this involves both angular and translational average)
and the integration over all metrics with the weight dictated by quantum
gravity.  Defined this way the exponential decay of $\la \Delta_g \ra$
should define the massive excitations and the short distance behaviour
should define the $\b$- and $\g$-functions,
i.e. the renormalization group equations.

Unfortunately the above program is rather inconvenient to implement
(see however \cite{david2} for a number of interesting observations). If we
only want the $\b$-function we can use the action itself. This
has the advantage that we work with quantities integrated over
space-time and we thereby avoid addressing explicitly the
concept of geodesic distances. Let us here discuss how to
obtain the $\b$-function for matter coupled
to gravity. The starting point is the simplest derivation of the
$\b$-function in the context of conformal field theory \cite{cardy}.
Let $S_0$ denote the action of the conformal invariant theory,
and let $V$ denote a marginal operator. The total theory will be
given by:
\beq{j1a}
S = S_0 + \l \int d^2 z V
\eeq
and will in general describe a perturbation away from the conformal
fixed point defined by $S_0$.
Denote the short distance cut-off $\ha$. The $\b$-function describes
the change in coupling constants needed to compensate a change in
cut-off for the partition function $Z(\ha,\l)$ or for the free energy
$F(\ha,\l) = -\log Z(\ha,\l) $:
\beq{j1b}
\left(- \ha \frac{\partial }{\partial \ha} +
\b(\l) \frac{\partial}{\partial \l}\right) F(\ha,\l) = 0.
\eeq
The free energy associated with $S$ can
be written as a power series in the coupling constant $\l$
and in this perturbation theory  the  dependence on the cut-off  $\ha$
due to ultra-violet divergences  can be found by assuming an
operator product  expansion. Let us for simplicity of the argument
assume that $V$ has the operator product expansion $V \cdot V = [V]$,
i.e.
\beq{j2a}
V(r) V(0) \sim \frac{c}{r^2} V(0),~~~~r \to 0.
\eeq
A term like
\beq{j3a}
\l^n\int d^2 z_1 \cdots d^2 z_n
\la V (z_1) \cdots V (z_n) \ra_{S_0}
\eeq
will have logarithmic singularities at coinciding arguments
$z_i \to z_j$. Using \rf{j2a} we get
\beq{j4a}
\int d^2 z_1 d^2z_2 \la \cdots V (z_1) V (z_2) \cdots\ra_{S_0} \sim
-2\pi c\cdot \log \ha  \int d^2 z \la \cdots  V (z) \cdots \ra_{S_0}.
\eeq
A change in cut-off $\ha \to \ha(1+dl)$ can be compensated by
a change $\l \to \l -dl  \pi c \l^2$ in the lower
order term in the perturbation expansion and this leads to the $\b$-function
\beq{j5a}
\b (\l) = -\frac{d\l}{dl} = \pi c\l^2+O(\l^3).
\eeq
The argument is easily generalized to the situation where the
operators $V$ are almost marginal. If the scaling dimension
of $V$ is assumed to be $2-y$, and we still insist that the
coupling constant $\l$ is  dimensionless, there will be an explicit
cut-off  dependence in the interaction term in \rf{j1a}:
\beq{j6b}
\frac{\l}{\ha^{y}} \int d^2 z V,
\eeq
and \rf{j5a} is generalized to
\beq{j5b}
\b (\l) = -\frac{d\l}{dl} = -y \l +\pi c\l^2 + O(\l^3).
\eeq

Let us consider the situation where the theory is coupled to
quantum gravity. We work in conformal gauge.
At the conformal point the coupled theory
is described by DDK. The metric is decomposed in a fiducial background
metric and the Liouville field by
\beq{j6}
g_{\a\b} = \hat{g}_{\a\b} e^{2\rho}.
\eeq
The Liouville part of the theory can be written as
\beq{j7}
S_L =
\frac{1}{8\pi} \int d^z \sqrt{\hg} \left(\hg^{\a\b}\prt_\a\rho\prt_\b \rho+
Q\hat{R}\rho+\m e^{\g \rho}\right),
\eeq
where
\beq{j7a}
 Q=\sqrt{\frac{25-c}{3}},~~~~~~\g=\oh(Q-\sqrt{Q^2-8})=
 \sqrt{\frac{25-c}{12}}-\sqrt{\frac{1-c}{12}}.
\eeq
In \rf{j7} the ghost part is left out as it will play no role in
the discussion to follow,
and in the matter part of $S_0$ $g_{\a\b}$ should be
replaced by $\hg_{\a\b}$.
Since we will be interested in the ultra-violet
properties of the theory we will presently work with the cosmological
term equal zero in accordance with the general folklore that this term
should play no role in the ultra-violet regime \cite{pol}.
We have two ultra-violet cut-offs: $\ha$ defined
in terms of the fiducial metric $\hg_{\a\b}$ and the physical cut-off $a$
defined by
\beq{j8}
ds^2 = e^{\g \rho} \hg_{\a\b}dz^\a dz^\b > a^2.
\eeq
The theory must be independent of the cut-off $\ha$ since the fiducial
metric is arbitrary: the $\b$-function must vanish.
However, if the matter theory defined  by \rf{j1a} before coupling to
gravity had a non-vanishing $\b$-function the arguments given above
can be repeated since the matter fields only couple to
the fiducial metric to lowest order and we have a dependence of $\ha$.
This dependence can be eliminated by including new couplings
between the Liouville field and the matter fields and it is a natural
conjecture that the interaction between matter and gravity
is uniquely fixed as a power series in the
coupling constant $\l$ by the requirement that all cut-off dependence of
$\ha$ cancels. To second order in $\l$ it is easy to determine
the coupling which leads to a vanishing $\b$-function. A glance on
\rf{j5b} shows that the $\b$-function will vanish if the operator
$\int d^2 z \sqrt{\hg} V$ acquired a scaling dimension $y$ such that
\beq{j8a}
 y (\l) = \pi c\l.
\eeq
While this makes no sense before coupling to gravity it is trivial to
implement after coupling to gravity since  the scaling dimension

\beq{j8c}
(\bar{\Delta}+\Delta) \int e^{y \rho/Q} V =-\frac{y}{Q}
\left( \frac{y}{Q} - Q\right) \;\approx y + O(y^2).
\eeq
If we expand in powers of $\l$ we get:
\beq{j9a}
\delta S = \frac{\pi}{Q} c \l^2 \int \rho V.
\eeq

The generalization to the more realistic situation where
the operator algebra is not as simple as \rf{j2a} is straight
forward. Let the perturbation away from the conformal
point be given by:
\beq{j1}
S = S_0 + \l_i \int d^2 z V_i,
\eeq
where summation over repeated indices is understood.
The operator product expansion is assumed to be
\beq{j2}
V_i(r) V_j(0) =  \frac{c_{ijk}}{r^2} V_k (0) +\cdots,
\eeq
and the corresponding $\b$-functions will be given by:
\beq{j5}
\b_k (\l) = -\frac{d\l_k}{dl} = \pi c_{ijk}\l_i\l_j + O(\l^3).
\eeq
After coupling to gravity the following change in the action will
ensure the vanishing of the $\b$-function to order $O(\l^3)$:
\beq{j9}
\delta S = \frac{\pi}{Q} c_{ijk}\l_i\l_j \int \rho V_k,
\eeq
a result first derived in \cite{schmid}. For an earlier approach to
the coupling of non-conformal matter to 2d gravity see \cite{tseytlin1}.

The results can  be generalized to non-linear sigma models.
Let us again restrict ourselves  to the simplest case where  we have
a non-linear sigma model on a  symmetric space
(this includes the $O(N)$ non-linear sigma model and the principal chiral
model). Let the model
before coupling to gravity be given by
\beq{j10}
S= \frac{1}{8\pi \l} \int d^2 z G_{ij}(X) \prt_\a X^i \prt_\a X^j.
\eeq
The coupling to gravity just consist in replacing $d^2z$
by $d^2z \sqrt{g}$ and one $\prt_\a$ by $g^{\a\b}\prt_\b$
and in the conformal gauge any explicit dependence on the
Liouville field drops out.
If we expand around the Gaussian fixed point of the non-linear
sigma model we will get an infinite power series of marginal
perturbations, but thanks to the symmetry we know that the
divergences can all be absorbed in a redefinition of the
coupling constant
\beq{j11}
\l \to \l Z(\l),~~~ Z(\l) = 1 +  C\l\;\log \ha + O(\l^2),
\eeq
corresponding to a $\b$-function
\beq{j12}
\b(\l) = -C \l^2.
\eeq
In \rf{j11} and \rf{j12} $C$ depends on the symmetric space.
Had we been in $2+\ep$ dimensions we would have picked up a linear term in
\rf{j12} due to the explicit dependence on the cut-off $\ha$ in the
action, as follows from dimensional considerations:
\beq{j12a}
\b(\l) = \ep \l - C \l^2 + O(\l^3).
\eeq
After coupling to gravity we can have no dependence on the cut-off
$\ha$ of the fiducial metric. Again, as above, we can obtain
a $\b$-function which is zero to $O(\l^3)$ by changing the dimension
of the action via coupling to the Liouville field:
\beq{j12b}
S= \frac{1}{8\pi \l} \int d^2 z \;\sqrt{\hg}\;e^{y(\l)\rho/Q}
G_{ij}(X)\;\hg^{\a\b} \prt_\a X^i \prt_\b X^j.
\eeq
To lowest order in $y$ we have
\beq{j12c}
\b(\l) = -y(\l) \l -C \l^2 + O(\l^3) = O(\l^3)~~~~{\rm for} ~~~~~
y= -C \l
\eeq
and the back-reaction of gravity on the non-linear sigma model
will be (to lowest order):
\beq{j12d}
\delta S = -\frac{C}{Q} \frac{1}{8\pi}\int d^2z \sqrt{\hg}\; \rho
\;G_{ij}(X)\; \hg^{\a\b} \prt_\a X^i \prt_\b X^j.
\eeq

The considerations for the non-linear sigma models are of course
of a rather formal nature since the central charge $c > 1$.

\subsection{The modified $\b$-function}

While the $\b$-function in terms of the unphysical cut-off
$\ha$ by construction is zero, the change in action as a result
of a change in the physical cut-off $a$ defined by
\rf{j8} leads to a universal modification of the original
$\b$-function \rf{j5}. The modification can be viewed as
the ``gravitational dressing'' of the $\b$ function.
Especially when formulated in the dynamical triangulated
approach \cite{david1,adf,kkm} it is clear that it is
the change in $a$ which governs the approach to the critical point.
The renormalized theory of Liouville and matter fields
was defined with respect to the fiducial cut-off $\ha$.
The physical cut-off $a$ appears in this formulation
rather indirectly as a lower bound on the Liouville
field $\rho$ through \rf{j8}:
\beq{j13a}
e^{\g \rho} \hg_{\a\b} dz^\a dz^\b \geq a^2 \Rightarrow
\rho \geq \rho_{min} = \frac{2}{\g} \log \frac{a}{\hat{r}}
\eeq
where $\hat{r}$ is some infrared cut-off defined in the
fiducial metric $\hg$.
Consider a change $a \to a(1+dl)$ of the physical cut-off. According
to \rf{j8} this leads to a change in the minimal  allowed
value of the Liouville field:
\beq{j13}
 \rho_{min} (a(1+dl)) \approx
\rho_{min} (a) + \frac{2}{\g} dl.
\eeq
This shift can be be viewed as a shift $\rho \to \rho +2dl/\g$
while keeping $\rho_{min}$ fixed. The ``running'' of the coupling
constants now appears by absorbing the shift $\rho \to  \rho+2dl/\g$
in a redefinition of the coupling constants $\l$  \cite{cst,schmid}.
To $O(\l^3)$ the only dependence on $\rho$ in the matter
part of the renormalized action
is found in $\delta S$ in \rf{j9} (or \rf{j12d}). A shift
$\rho \to  \rho+2dl/\g$ leads to a  change of $\delta S$ by
\beq{j14}
dl\;\frac{2}{\g Q} \; c_{ijk} \l_i\l_j \int V_k,
\eeq
and it can be absorbed to order $O(\l^3)$ by a change in coupling constants:
\beq{j15}
\l_k \to \l_k - dl \frac{2}{\g Q} c_{ijk} \l_i \l_j
\eeq
The change in coupling constants needed to absorb the change in
$a$ is thus the same as before coupling to gravity, except for a
factor $2/\g Q$ and
the modified $\b$-function, $\b_G$, defined as $-d\l/dl$,  is related
to $\b$-function before coupling to gravity as
\beq{j16}
\b_G (\l) = \frac{2}{\g Q} \b (\l).
\eeq
This result is (implicitly) in the more general discussion in \cite{schmid}
and was rediscovered in \cite{5} and we see that it is also
(at a formal level) valid for non-linear sigma models.
We note that the sign of
the $\b$ function will not be changed by the coupling to gravity
as long as $\g$ and $Q$ are positive, i.e as long as $ c \leq 1$.
In case $c >1$ $\g$ becomes complex and we encounter of course
in this region the well known problems of coupling of matter to
2d gravity. Note also that the correction  is 1/2 for $c=1$ and
decreases as $c \to -\infty$ where $2/\g Q \to 1$. This is in accordance
with the idea that the fluctuating geometries are suppressed for
$c \to -\infty$. Indeed, the fractal dimension of space-time as
defined in \cite{kn} is given by
\beq{j17}
D_F = \left( 1 + \sqrt{\frac{13-c}{25-c}}\right) \frac{\g Q}{2},
\eeq
which goes to 2 for $c \to -\infty$.
Since the $\b$-function in a fixed geometry determines the
change of coupling constants with distance it is interesting to
try to understand whether such an interpretation can be given
after coupling to gravity. It should involve a proper treatment
of the quantum average of geodesic distances as discussed above.

\section{The effective action in target space}

Consider the following action,
\beq{1}
S_{n,\eta}={1 \over 4\pi}\int\,d^2z\sqrt{g}
\left\{-\eta\phi^n R+
{1 \over 2}\prt_\a\phi\prt^\a \phi+V(\phi)
            +{1 \over 2}\sum_{i=1}^N \prt_\a f^i \prt^\a f^i\right\}.
\eeq
The matter content of the theory is $N$ scalar fields with a Gaussian
interaction and one scalar field $\phi$ which has non-trivial self-interaction
via the potential  $V(\phi)$. The coupling to gravity is  inherent in the
area element $d^2 z \sqrt{g}$ and the derivatives
$\prt^\a = g^{\a\b} \prt_\b$. The cosmological term is
contained in the constant term of the potential $V(\phi)$.
In addition we have introduced an explicit coupling between
$\phi$ and the curvature. We will consider here only the cases $n=0$ and
$n=1$. It would be interesting to be able to consider higher $n$'s
in analogy with
the non-minimal coupling of a scalar field to gravity in higher dimensions.
However, in these cases we have a non-trivial interaction and
we only know how to include the term as a perturbation in $\eta$, which
is not what we want. The Gaussian matter fields $f^i$ will couple
to gravity only by the conformal anomaly, i.e. only through
their central charge $c=N$ and we have only included
them in order to be able to vary the total central charge of the matter
sector.  The model \rf{1}
has been considered in various contexts \cite{mann,mann1}.
Let us here review the part we need for the target space formulation.

Let us work in conformal gauge by choosing
a fiducial metric ${\hat g}_{\a\b}$ as
\beq{2}
g_{\a\b}=e^{2\rho}{\hat g}_{\a\b},
\eeq
\noindent where $\rho$ represents the conformal mode.
The quantum measures of $\rho,\phi$ and $f^i$ are introduced
according to DDK, and the effective action can be
written as
\bea
S_{n,\eta}={1 \over 4\pi}\int\,d^2z\sqrt{{\hat g}}
           & \left\{
              -2n\eta\phi^{n-1}{\prt_\a}\phi{\prt^\a}\rho
            +{1 \over 2}\prt_\a\phi\prt^\a\phi
            +{\kappa \over 2}{\prt_\a}\rho\prt^\a \rho\right. \nn
           &\left.-{\hat R}(\eta\phi^n-{1 \over 2}\kappa\rho)
            +{\hat V}(\phi,\rho)\right\}
            +S_{f}+S_{ghost}, \label{3}
\eea
\noindent where
$$ \kappa={24-N \over 3}, $$
\noindent and $\prt^\a= \hg^{\a\b}\prt_\b$ and ${\hat R}$ is
the scalar curvature for the fiducial metric
${\hat g}_{\a\b}$.
$S_{ghost}$ and $S_f$ play no role in the arguments to follow
and we will ignore them. It is convenient to rescale $\rho \to
\rho\sqrt{\kappa}$ to normalize the kinetic term. After this
modification the effective action is:
\bea
S_{n,\eta}={1 \over 4\pi}\int\,d^2z\sqrt{{\hat g}}
           & \left\{
              -\frac{2n\eta}{\sqrt{\kappa}}\;
              \phi^{n-1}{\prt_\a}\phi{\prt^\a}\rho
            +{1 \over 2}({\prt}\phi)^2
            +{1 \over 2}({\prt}\rho)^2\right. \nonumber\\
           &\left.-{\hat R}(\eta\phi^n-{1 \over 2}\sqrt{\kappa}\rho)
            +{\hat V}(\phi,\rho/\sqrt{\kappa})\right\},  \label{3'}
\eea

Since \rf{1} is invariant under the transformation,
$$ {\hat g}_{\a\b}\to {\hat g}_{\a\b}e^{\sigma (z)},
        \qquad \rho\to\rho-{\sigma \over 2}, $$
and since the DDK measure $\cD \rho$ is invariant under translations,
the partition function $Z({\hat g})$, which is obtained
by integrating over $\rho$, $\phi$, $f^i$ and ghosts, must be
invariant under the conformal transformation
\beq{4}
{\hat g}_{\a\b}\to {\hat g}_{\a\b}e^{\sigma (z)}.
\eeq
Eq.\rf{3} (or \rf{3'})  is invariant under the above conformal transformation
if ${\hat V}$ is zero and $n \leq 1$ (we consider only these choices of
$n$). For a general potential the task is to find the modification
$V(\phi) \to {\hat V}(\phi,\rho)$ consistent with the conformal invariance
\rf{4}.

In the approach of DDK, ${\hat V}$ has been determined by the requirement
that the  term  $\int {\hat V}$ by itself should be invariant under
the conformal transformations \rf{4} if calculated around the conformal point
where $V = 0$, i.e. by imposing that ${\hat V}(\phi,\rho)$ is a $(1,1)$
operator:
$$(L_0+{\bar L}_0){\hat V}=2{\hat V}.$$
If $V$ itself has definite scaling properties a solution to this
equation can be found in the form ${\hat V}(\phi,\rho) = V(\phi)e^{\a \rho}$.
However, writing $V = \l P$, we saw in the last section that even if
$V$ was a marginal perturbation this result will only be the lowest
order solution in $\l$ if the $\b$-function was different from zero.
The same situation appears in dilaton gravity and it is an unsolved
problem how to obtain the effective action which includes the gravitational
effects to all orders.  Here we will confine ourselves to study the problem
perturbatively in the coupling constant $\l$ of the potential $V$.

Eq. \rf{3'} can be rewritten as a general non-linear sigma model:
\beq{5} S_{eff}=
{1 \over 4\pi}\int\,d^2z\sqrt{{\hat g}}
            \left[ {1 \over 2}G_{\mu\nu}(X){\hat g}^{\a\b}
              \partial_\a X^{\mu}\partial_\b X^{\nu}
           +{\hat R}\Phi(X)+T(X) \right],
\eeq
where $X^{\mu}=(X^0,X^1)=(\rho,\phi)$. In the terminology
of string theory, $G_{\mu\nu}(X)$, $\Phi(X)$ and $T(X)$ represent
the metric, dilaton and tachyon in the target space, respectively.
They can be determined by solving the equations of zero $\beta$ functions
of the non-linear $\sigma$ model eq. \rf{5}. As explained above
the $\b$-functions
should be zero  since the partition function, as a function of $\hg$
is conformally invariant. The solutions $G_{\m\n}(X)$, $\Phi (X)$ and
$T(X)$ obtained from the vanishing of the $\b$-functions of \rf{5}
can be gotten as solutions to
the classical equations of motion of the following
effective action in target space \cite{cst,4}:
\beq{6}
S_t=
{1 \over 4\pi}\int\,d^2X\sqrt{G} e^{-2\Phi}
            \bigl[ R-4(\nabla\Phi)^2+
                  {1 \over 16}(\nabla T)^2+{1 \over 16}v(T)+\kappa
            \bigr],
\eeq
\noindent where
\beq{7}
v(T)=-2T^2+{1 \over 6}T^3+\cdots
\eeq
and where $\nabla_\m$ denotes the covariant differentiation with
respect to the metric $G_{\m\n}$.
The higher derivative terms are suppressed, and they will
be necessary in a systematic higher order expansion.
As for the powers of $T(X)$, it
may be possible to remove the higher order terms in $v(T)$
(from $O(T^3)$) by the field redefinition in the target space \cite{7}.
However we want to
keep the relation between \rf{3} and \rf{6}, so we adopt $v(T)$ of eq. \rf{7}.

\vspace{12pt}

The equations of motion derived from the above action are the coupled
equations of $T$ and other fields, and in this way we see in a very
direct way the corrections needed for the potential $V(\phi)$. This
is our  main motivation for using the reformulation of
the problem of gravitational dressing as a set of classical
equations of motion. In addition it allows us to address the question
of gravitational dressing for perturbations which are not necessary
marginal, thereby generalizing the treatment in sect. 2.

\vspace{12pt}

{}From \rf{6} we obtain the classical equations of motions for
$T(X)$, $\Phi(X)$ and $G_{\mu\nu}(X)$:
\bea
\nabla^2T-2\nabla\Phi \nabla T&=&{1 \over 2}v'(T),\label{8} \\
 \nabla^2\Phi-2(\nabla\Phi)^2&=&{\kappa \over 2}
                       +{1 \over 32}v(T), \label{9} \\
 R_{\mu\nu}-{1 \over 2}G_{\mu\nu}R &=&
            -2\nabla_{\mu}\nabla_{\nu} \Phi+G_{\mu\nu}\nabla^2\Phi
            +{1 \over 16}\nabla_{\mu}T\nabla_{\nu}T-{1 \over 32}G_{\mu\nu}
            (\nabla T)^2. \label{10}
\eea
The procedure for solving \rf{8}-\rf{10} is as follows. Consider the
potential of the form
$$ V(\phi)=\lambda P(\phi), $$
\noindent where $\lambda$ is a small parameter. We expand
$G_{\mu\nu}$, $\Phi$ and $T$ as
\bea
G_{\mu\nu}&=&
        \pmatrix{1 & -\frac{2n\eta}{\sqrt{\kappa}}\phi^{n-1}\cr -
        \frac{2n\eta}{\sqrt{\kappa}}\phi^{n-1}&1}
        +\lambda^2h_{\mu\nu}+\cdots , \label{11} \\
\Phi&=&\Phi^{(0)}+\lambda^2\Phi^{(2)}+\cdots, \label{12} \\
T&=&\lambda(T^{(0)} +\lambda T^{(1)}+\cdots),   \label{13}
\eea
where $Q=\sqrt{\kappa}$ and $\Phi^{(0)}={1 \over 2}Q\rho-\eta\phi^n$.
{}From eqs.\rf{7}-\rf{10} it follows that the lowest order
corrections to $G_{\mu\nu}$ and $\Phi$ are of order $O(\lambda^2)$.
They are chosen as to reproduce the original action eq.\rf{3}
with ${\hat V}(\rho,\phi)=0$ in the limit $\lambda \to 0$. \par
We can now start solving eqs.\rf{8}-\rf{10} perturbatively with respect to
$\lambda$ order by order. The first step is to solve
the lowest order equation which
comes from \rf{8}. This determines  $T^{(0)}$ and
is equivalent the method of DDK.

Explicitly the lowest order equation is (for $\eta = 0$):
\beq{ad1}
\partial^2T^{(0)}-2\partial\Phi^{(0)}\cdot\partial T^{(0)} =-2T^{(0)},
{}~~~~\Phi^{(0)} =\oh Q\rho.
\eeq
If the potential $V(\phi)$ has definite scaling properties
the lowest order form of $T^{(0)}$ is the DDK ansatz $T^{(0)} =
V(\phi) e^{\a \rho}$. This follows since the only functions
of $\phi$ with definite scaling properties are exponentials,
and if we use exponentials like $e^{ip\phi}$ (or better $\cos p\phi$ and
$\sin p\phi$) in \rf{ad1} we get the DDK result:
 \beq{15}
 \alpha={1 \over 2}\biggl[Q-\sqrt{Q^2+4p^2-8}\biggr].
\eeq
However, since the lowest order equation \rf{ad1} is linear  the
solution can be found for any potential by Fourier transformation.
The result is
\beq{ad2}
T^{(0)} (\phi,\rho) = e^{\oh Q\rho}
\int d\xi \;P(\phi-\xi) \;\frac{4\pi q \rho}{\sqrt{\rho^2+\xi^2}}
K_1 (q \sqrt{\rho^2+\xi^2}),~~~~q=\sqrt{\frac{1-c}{12}}.
\eeq
For $N=0$ we have $c=1$, $\g=Q/2=\sqrt{2}$
and the expression simplifies a little:
\beq{ad3}
 T^{(0)} (\phi,\rho) =  e^{\g\rho}
\int d\xi P(\phi-\xi) \frac{4\pi \rho}{\rho^2+\xi^2}.
\eeq

We can then  solve the next order equations by using \rf{ad2} or \rf{ad3}
in eqs.\rf{8}-\rf{13}. We  illustrate the procedure in two simple models
corresponding to $n=0$ and $n=1$. \par

\section{ n=0: The Sine-Gordon model}

For $n=0$, the first term in eq.\rf{1} is a total divergence, so we neglect
it here. As for the potential we consider the  Sine-Gordon model,
\beq{14} V(\phi)=\lambda\cos(p\phi).
\eeq
The choice of this potential is dictated by our desire
have a simple expression  for $T^{(0)}$ as given by \rf{ad2} or \rf{ad3}.
If we take $N=0$ we get\footnote{For a different treatment of $T^{(0)}$ see
\cite{tseytlin}.} from the above formulas:
\beq{ad4}
T^{(0)}= \cos p\phi \;e^{\alpha\rho},~~~~~~
 \alpha=\sqrt{2} -p.
\eeq
Using this result, we obtain the next order of \rf{8}-\rf{10} ($O(\lambda^2)$)
by expanding according to eqs.\rf{11} and \rf{12} (with $\eta =0$)
\beq{16}
{4 \over Q}(\partial_1^2-\partial_0^2)\Phi^{(2)}+
\partial_0(h_{00}+h_{11})-2\partial_1h_{10} =
-{1 \over 16Q}\bigl[(\alpha^2+p^2)\cos(2p\phi)
              +\alpha^2-p^2\bigr]e^{2\alpha p}
\eeq
\beq{17}
-{4 \over Q}\partial_1\partial_0\Phi^{(2)}+
   \partial_1 h_{00}={\alpha p \over 16Q}\sin (2p\phi)
                       e^{2\alpha\rho},\hspace{2cm}
\eeq
\beq{18}
{4 \over Q}(\partial^2-2Q\partial_0)\Phi^{(2)}+
\partial_0h_{11}-2\partial_1h_{10} =
-2Qh_{00}+\partial_0 h_{00}
            -{2 \over 16Q}[\cos (2p\phi)+1]e^{2\alpha\rho}
\eeq
where $\partial_{0,1}$ mean the partial derivative with
respect to $\rho$ and $\phi$, respectively.

We note the following. Since we solve for the higher order terms
$h_{\mu\nu}$, $\Phi^{(2)}$ and $T^{(1)}$ as the response
of the lowest order terms $T^{(0)}$, $\Phi^{(0)}$ we should only
use the particular solutions which go to zero with $T^{(0)}$.
Next we will expand around the point $(p,\a) =(\sqrt{2},0)$.
$p = \sqrt{2}$  is the Kosterlitz-Thouless transition point for the
Sine-Gordon equation before coupling to gravity, and exactly at this
point $V(\phi) =\l \cos p\phi$ is a marginal perturbation. From
sect. 2 it follows that a sensible ansatz for $h_{\m\n}$ is
\beq{20}
 h_{\mu\nu}=\left(\matrix{
          0 & 0 & \cr
          0 &  h(\rho) & }   \right).
\eeq
In this case, the eqs.\rf{16}-\rf{18} can be written as,
\bea
{4 \over Q}(\partial_1^2-\partial_0^2)\Phi^{(2)}+
\partial_0 h
            &=&-{1 \over 16Q}\bigl[(\alpha^2+p^2)\cos(2p\phi)
              +\alpha^2-p^2\bigr]e^{2\alpha\rho},\hspace{.5cm} \label{16'} \\
-{4 \over Q}\partial_1\partial_0\Phi^{(2)}
   &=&{\alpha p \over 16Q}\sin (2p\phi)
                       e^{2\alpha\rho}, \hspace{.5cm}\label{17'}\\
{4 \over Q}(\partial^2-2Q\partial_0)\Phi^{(2)}+
\partial_0 h&=&
            -{2 \over 16Q}[\cos (2p\phi)+1]e^{2\alpha\rho}.
            \hspace{.5cm}\label{18'}
\eea
{}From these equations, we get two equations for $\Phi^{(2)}$. One
is given by eq.\rf{17'} and the other is obtained by subtracting \rf{18'}
from \rf{16'}. The latter is written as,
\beq{21} (\partial_0^2-Q\partial_0)\Phi^{(2)}
            ={1 \over 128}\bigl[(2\alpha^2-Q\alpha)\cos(2p\phi)
              +Q\alpha-4\bigr]e^{2\alpha\rho}.
\eeq
\noindent Both eqs.\rf{17'} and \rf{21} can be solved for infinitesimal
$\alpha$ as follows,
\beq{22}
\Phi^{(2)}={1 \over 256}\cos(2p\phi)+{1 \over 32Q}\rho+O(\alpha).
\eeq
\noindent By substituting this into \rf{16'} (or \rf{18'}), we obtain
\beq{23}
h ={p^2 \over 16Q}\rho+O(\alpha).
\eeq

Finally, we solve for $T^{(1)}(X)$ by taking the
$T^3$ term in $v(T)$ into account. The equation reads:
\beq{26}
 (\partial^2-Q\partial_0+2)T_1={1 \over 8}
                (\cos 2p\phi+1)e^{2\alpha\rho},
\eeq
\noindent and its special solution near $\alpha=0$ is obtained as
\beq{27}
T_1={1 \over 32}(1-\cos 2p\phi)+O(\alpha),
\eeq
As a result, we obtain the
effective action up to $O(\lambda^2)$ near ($p=\sqrt{2}$, $\alpha=0$)
as follows,
$$ S_{0,eff}=S_0+\lambda^2S_0^{(2)}+\cdots, $$
\bea
 S_{0}&=&{1 \over 4\pi}\int\,d^2z\sqrt{{\hat g}}
            \biggl\{ {1 \over 2}(\partial\phi)^2 +{1 \over 2}(\partial\rho)^2
            +{\hat R}\Phi^{(0)}
+\lambda\cos(p\phi)e^{\alpha\rho}\biggr\} ,  \label{28}\\
&& \nn
 S_0^{(2)}&=&{1 \over 4\pi}\int\,d^2z\sqrt{{\hat g}}
            \biggl\{
            {1 \over 16Q}\rho (\partial\phi)^2
       +\bigl[{1 \over 256}\cos(2p\phi)+{1 \over 32Q}\rho\bigr]{\hat R}\nn
            &&\hspace{7cm}+{1 \over 16}(1-\cos(2p\phi))\biggr\} ,  \label{29}
\eea
\noindent where we have neglected the terms of order $O(\lambda^2\alpha)$
in $S_0^{(2)}$ and where $e^{\a \rho}$ should
be understood as expanded in powers of $\a$, i.e
(to lowest order) in  $\sqrt{2}-p$. \par

\vspace{12pt}

Let us now consider the  renormalization group equations of the
coupling constants  $p$ and $\lambda$. As explained in sect. 3
the change in coupling constants should be obtained
by absorbing a shift $\rho \to \rho +2dl/\g = \rho + \sqrt{2} dl$.
Let us write $p = \sqrt{2}+\ep$ and denote by $\l',\ep'$ and $\phi'$ the
coupling constants and fields after the shift of $\rho$ by $dl/2$.
We note that the first term in \rf{29} is responsible for a wave
function renormalization:
\beq{30}
{\phi}^2 = (1+ \frac{\l^2 }{8} \frac{2}{\g Q} dl){\phi'}^2
\eeq
Since $p \phi= p'\phi'$ we get
\beq{31}
\ep' =\ep  +\frac{\sqrt{2} \l^2}{16}\frac{2}{\g Q} dl,~~~~~~~
\l' = \l-\a \l\;\frac{2}{\g}dl.
\eeq
{}From \rf{ad4} we have $\a =\sqrt{2} -p = -\ep$ to order $O(\l^2)$,
and since (again to lowest order) $Q= 2\sqrt{2}$ and $\g = \sqrt{2}$,
we can write the renormalization group equations as
\beq{33} \frac{d\lambda}{dl}=\sqrt{2}\epsilon\lambda, \qquad
   \frac{d\epsilon}{dl}={1 \over 16\sqrt{2}}\lambda^2,
\eeq
If we compare these results with the renormalization group equations
obtained without the coupling to gravity  \cite{8},
\beq{33'}
\frac{d \lambda}{dl}=2\sqrt{2}\epsilon\lambda, \qquad
   \frac{d\epsilon}{dl}={1 \over 8\sqrt{2}}\lambda^2,
\eeq
we see that they differ by the factor $1/2=2/\g Q$ in agreement with
the general discussion in sect. 3. In addition we see that the term
$\rho  \hat{R}/32Q$ induces as change $Q \to Q+\l^2/32Q$.
It is interesting to note that an increase in $Q$ formally is
in accordance with a renormalization flow way from the ultra-violet
fixed point towards a new infrared stable fixed point with a smaller $c$,
as one would expect from the $c$-theorem.

\section{n=1: The extended Sine-Gordon model}

The model of $n=1$ without the kinetic term of $\phi$ was first
proposed in \cite{9} and examined
by several people as a model of the non-critical string beyond
$c=1$ \cite{10} due to the fact that the string susceptibility
remained real for any value
of $c$. However, such a model constrains the value of the scalar curvature
$R$ to such an extent that it ceases to be a dynamical variable.
Here we consider a model where a kinetic $\phi$ term is added
so $\phi$ is no longer
an auxiliary field and the scalar curvature a dynamical variable.

The analysis of cosmological
term gives a constraint on $N$ from
the reality of the dressed factor $e^{\g \rho}$.
This is nothing but the reality of the string
susceptibility. Consequently we first analyze the cosmological term.

\subsection{The cosmological term and the string susceptibility.}

We parametrize the cosmological term as,

$$ V_c={\bar \lambda}e^{\gamma_1\rho}. $$

\noindent By rescaling the variable and parameter in eq.\rf{3} as,
$Q\rho\to \rho$ and
$\eta/Q \to \eta$, we obtain the following result from the lowest order
equation for $T$.
\beq{35}
\gamma_1={1 \over 2}\biggl[Q{\bar \eta}-
            \sqrt{Q^2{\bar \eta}^2-8{\bar \eta}}\biggr],
\eeq
\noindent where ${\bar \eta}=1-4\eta^2$.

We demands that $\gamma_1$ is real and find \par
(i) if $\eta^2<{1 \over 4}$ (${\bar \eta}>0$) then  $ N\leq 0$,~~~ i.e.~~~
 $N=0$ \par
(ii) if $\eta^2>{1 \over 4}$ (${\bar \eta}<0$) then $N\leq 24$. \par
\noindent $c$ could exceed one in case (ii), but the theory is not
unitary since $\det G_{\mu\nu}=1-4\eta^2<0$.
If we respect  the unitarity of the two dimensional system,
we must restrict ourself to the case (i).
However, In order to see what happens in the somewhat ill-defined
(non-unitary) theory, we also examine the solutions in case (ii).\par
The string susceptibility, which is denoted by ${\hat\gamma}$
can be calculated by the method of DDK and we get
\beq{36}
{\hat\gamma}=2-{\chi \over 24}\biggl[
             24-N+\sqrt{(24-N)(24-N-{24 \over {\bar\eta}})}\biggr],
\eeq
\noindent where $\chi={1 \over 4\pi}\int d^2z\sqrt{{\hat g}}{\hat R}.$
${\hat \gamma}$ is real if $\gamma_1$
is real. If we consider the case (ii), ${\hat \gamma}$ is real even
if c is larger than one, but as noted above unitarity is broken.
In this sense we can not exceed the $c=1$ barrier.

\subsection{  $ \eta^2<{1 \over 4}$}

 Since the reflection symmetry, $\phi\to -\phi$, is broken
due to the term $\phi R$ in this model, we extend  $V(\phi)$ as
\beq{37}
V(\phi)=\lambda[\cos(p\phi)+\delta\sin(p\phi)]
\eeq
\noindent by adding the odd term, $\sin(p\phi)$, with a weight $\delta$.
Even if we start from the even term $V(\phi)=\lambda\cos(p\phi)$, we will
find that the odd term $\sin(p\phi)$ is needed in order to solve the lowest
order of equation \rf{8}. So it is possible to consider the odd term as
a correction term due to quantum gravity. In any case,
we can take the form eq.\rf{37} with a dressed factor as the lowest order
solution of eq.\rf{8}. \par
 For the simplicity, we further change and rescale variables in
$I_{1,\eta}$ after the change, $\sqrt{\kappa}\rho\to\rho$, as follows

$$\rho\to \rho+2\eta\phi, \qquad \phi^2{\bar\eta}\to\phi^2. $$

\noindent Since the Jacobian is a simple constant,
there is no problem with this change of variables.
The action is written as
\beq{38}
S_{1,\eta}={1 \over 4\pi}\int\,d^2z\sqrt{{\hat g}}
            \bigl\{ +{1 \over 2}({\prt}\phi)^2
            +{1 \over 2}({\prt}\rho)^2
           +{\hat R}{Q \over 2}\rho
            +{\hat V}(\phi,\rho+{2\eta \over \sqrt{\bar\eta}}\phi)
            \bigr\},
\eeq
We have to work with the dressed operator
$e^{\alpha(\rho+{2\eta \over \sqrt{\bar\eta}}\phi)}$ instead of
$e^{\alpha\rho}$. Except for this point, the analysis is
parallel to the one in the previous section.
In particular we note that since $N=0$ we have $c=1$ for
$\l=\delta =0$. This means that $Q=2\sqrt{2}$ and $\g=\sqrt{2}$
and that we    have to work close to
the Kosterlitz-Thouless transition point $p = \sqrt{2}$ in order
that the perturbations are almost marginal and we can use the ansatz
\rf{20}.

First we solve the lowest order equation of $T^{(0)}$ in the form,
\beq{40}
T^{(0)}=[\cos(p\phi)+\delta\sin(p\phi)]
                 e^{\alpha\rho+\beta\phi},
\eeq
\noindent where
 $$\beta= {2\eta \over \sqrt{1-4\eta^2}}\alpha.  $$

Since $T^{(0)}$ contains two independent functions, $\sin(p\phi)$
and $\cos(p\phi)$ we obtain two equations of the parameters from the
lowest order equation of $T^{(0)}$,
\bea
& \alpha^2+ \beta^2-p^2+2\beta p\delta-Q\alpha+2=0,& \label{41}\\
&   \delta(\alpha^2+\beta^2-p^2)-2\beta p-\delta(Q\alpha-2)=0.\label{42}
\eea
{}From these equations, we obtain $\alpha=0=\beta$ for $p\neq 0$.
Since we  consider the case of $p^2=2$ we have that $\alpha=0$. Near this
point, the solution of the next order equations with ansatz eq.\rf{20}
are as follows,
\bea
 h&=&{p^2 \over 16Q}(1+\delta^2)\rho+O(\alpha). \label{43}\\
 \Phi^{(2)}&=&{1 \over 32Q}(1+\delta^2)\rho+O(\alpha).\label{44}\\
T^{(1)}&=& -{1-\delta^2 \over 48}\cos(2p\phi)
  -{\delta \over 24}\sin(2p\phi)+{1+\delta^2 \over 16}+O(\alpha).\label{45}
\eea
We can obtain the renormalization group equations for the
coupling constants $\l$, $\ep$ ($=p-\sqrt{2}$) and $\eta$ similarly
to the case in the last section:
\beq{46}
 {\dot \lambda}=\sqrt{2}\epsilon\lambda, \qquad
   {\dot \epsilon}={1+\delta^2 \over 16\sqrt{2}}\lambda^2,\qquad
   {\dot \eta}=-{1+\delta^2 \over 16\sqrt{2}\epsilon}\eta\lambda^2,
\eeq
It is  possible to compare with the results
obtained in \cite{11}. Again we see the difference of a factor
two of the coefficients, compared to model without coupling to
gravity. In addition eq.\rf{44} gives a shift of the charge $Q$ by
${\lambda^2 \over 32Q}(1+\delta^2)$. This is similar to
the case of $n=0$, and the same remarks apply here.

\subsection{  $\eta^2>{1 \over 4}$, Non-unitary Model }

In this case we can repeat the analysis by making the replacement
$ \phi^2{\hat\eta}\to\phi^2,$
where ${\hat \eta}=4\eta^2-1(>0)$. Then the action can be  written as,
\beq{49}
S_{1,\eta}={1 \over 4\pi}\int\,d^2z\sqrt{{\hat g}}
            \left\{ -{1 \over 2}({\prt}\phi)^2
            +{1 \over 2}({\prt}\rho)^2 \nn
           +{\hat R}{Q \over 2}\rho
            +{\hat V}(\phi,\rho+{2\eta \over \sqrt{\hat\eta}}\phi)
            \right\},
\eeq
and we must use the following lowest order metric in the target
space,
$$ G_{\mu\nu}^{(0)}=\left(\matrix{
          1 & 0 & \cr
          0 &  -1  & }   \right).  $$
By using the same form of $T^{(0)}$ as in \rf{40}
with $\beta={2\eta \over \sqrt{4\eta^2-1}}\alpha$, we obtain the
following results from the lowest order equation \rf{8},
\bea
&\alpha^2-\beta^2+p^2-2\beta p\delta-Q\alpha+2=0,& \label{50} \\
&\delta(\alpha^2-\beta^2+p^2)+2\beta p-\delta(Q\alpha-2)=0.& \label{51}
\eea
\noindent These equations have {\it no solution} at $p^2=2$, and we obtain,
\beq{52}
p=0, \qquad\qquad \alpha^2-\beta^2-Q\alpha=-2.
\eeq
The consistent solution is then
$$h_{\mu\nu}=0,$$
and we can write three equations coming from eqs.\rf{9}
and \rf{10} as
\bea
 (\partial_1^2+\partial_0^2)\Phi^{(2)} &=&{1 \over 32}(\alpha^2+\beta^2)
               e^{2(\alpha\rho+\beta\phi)}, \label{54}\\
  \partial_1\partial_0\Phi^{(2)} &=&{\alpha\beta \over 32}
                       e^{2(\alpha\rho+\beta\phi)}, \label{55}\\
(\partial^2-2Q\partial_0)\Phi^{(2)}
            &=&-{1 \over 16}e^{2(\alpha\rho+\beta\phi)}, \label{56}
\eea
They admit the solution
\beq{57}
\Phi^{(2)}={1 \over 128}e^{2(\alpha\rho+\beta\phi)}.
\eeq
\noindent Further, from the next order of eq.\rf{8}, we obtain
\beq{58}
T^{(1)}=-{1 \over 16}e^{2(\alpha\rho+\beta\phi)}.
\eeq
In this case the critical value of $\alpha$ depends of $N$ which
could exceed zero but was bounded by
\beq{59}
N<12\beta^2,
\eeq
at the price of unitarity. Further $\alpha$ to lowest order is given by
\beq{60}
\alpha={1 \over 2}\biggl[-Q{\hat \eta}+
            \sqrt{Q^2{\hat \eta}^2+8{\hat \eta}}\biggr] = \g_1.
\eeq
where $\g_1$ denotes the exponential factor in the cosmological term.
We can formally write down the renormalization group equation
for $\lambda$ near some value of $\alpha$ as,
\beq{61}
{\dot \lambda}=-\alpha^2\lambda.
\eeq
The important difference between the unitary and non-unitary
models is that we can not choose $p^2=2$ in the non-unitary case and
the equation for $\alpha$ can not be obtained in this case.

\section{The non-linear sigma models}

Let us finally discuss the target space equations for the last
class of models discussed in sec. 2, the non-linear sigma models.
As already remarked in sec. 2 the discussion is somewhat formal
in the sense that these models in general will have $c > 1$. To
simplify the discussion we consider here the simplest model, the
$O(N)$ non-linear sigma model. When we couple the model to gravity
by DDK quantization we get (ignoring as usual the ghost terms)
\beq{sgm1}
S(\hg,\l) = \frac{1}{8\pi} \int d^2 z\sqrt{\hg}
\left\{ \prt_\b \rho \prt^\b \rho +Q \hat{R}\rho\right\}
 +\frac{1}{8\pi\l}\int d^2 z\sqrt{\hg}\; e^{\a\rho}  \sum_{i=1}^N
\prt_\b \phi^i \prt^\b \phi^i,
\eeq
where $\sum_{i=1}^N \phi^i\phi^i =1$, and where
$\rho$ as usual denotes the Liouville field.
We have inserted the factor $e^{\a\rho}$ which we in sec. 2
argued should be present when we move away from the ultra-violet fixed
point corresponding to $\l =0$. We further expect $\a = O(\l)$ and
can a priori only expect this interaction between the Liouville
field and matter fields to be correct to this order. From this
point of view it is a little misleading to write the interaction
in the exponential form.

If we expand around $ \phi^N = \sqrt{1-\sum_{i=1}^{N-1} {\phi^i}\phi^i}$,
and rescale $\phi \to \sqrt{\l} \phi$ \rf{sgm1} can be written as
\beq{sgm2}
S(\hg, \l) = \frac{1}{8\pi}
\int d^2 z \sqrt{\hg} \left\{  \prt_\b \rho \prt^\b \rho
+ Q\hat{R}\rho+ \sum_{i=1}^{N-1} (\prt_\b \phi^i \prt^\b \phi^i) +
\hat{V}(\rho,\phi) \right\},
\eeq
where the potential $\hat{V}$ is given by
\bea
\hat{V} &=& \sum_{i,j}^{N-1}\;\left\{
\frac{\l\phi^i\phi^j}{1-\l\sum_k^{N-1} \phi_k^2}+
(e^{\a\rho}-1)\dl_{ij} \right\}\; \prt_\b \phi^i \prt^\b \phi^j \nonumber\\
&=& \sum_{i,j}^{N-1} (\l \phi^i \phi^j + \dl_{ij}\a \rho)
\prt_\b \phi^i \prt^\b \phi^j +O(\l^2,\a^2,\l\a).  \label{sgm3}
\eea
With the chosen expansion the indices $\m$ and $\n$ in target
space runs from 0 to $N-1$,
where 0 corresponds to the Liouville field $\rho$ and the vector
$X^\m$ defined below \rf{5} will in this case by written as
$X^\m = (X^0,X^i) = (\rho, \phi^i)$. The expansion of the target space
metric $G_{\m\n}(X)$,  dilaton $\Phi (X)$ and tachyon $T(X)$ will
in analogy  with \rf{11}-\rf{13} be given by
\bea
G_{\mu\nu}&=&G^{(0)}_{\m\n} + \l h_{\m\n}
        +\cdots , \label{sgm4} \\
\Phi&=&\Phi^{(0)}+\lambda\Phi^{(1)}+\cdots, \label{sgm5} \\
T&=& T^{(0)} +\lambda T^{(1)}+\cdots,   \label{sgm6}
\eea
where $T^{(0)} =0$, $\Phi^{(0)} = Q\rho/2$,  and
\beq{sgm7}
G^{(0)}_{\m\n} = \dl_{\m\n},~~~~~~~
h_{\m\n} = \pmatrix{0 & 0\cr 0 & \phi^i\phi^j +A \dl_{ij}\rho}.
\eeq
Since we expect that $\a = O(\l)$ we have in \rf{sgm7} introduced
the notation
$$\a = A \l + O(\l^2)$$.

{}From \rf{9} and\rf{10} we get to lowest order after a some calculations:
\beq{sgm8}
\prt^2\Phi^{(1)} -2 Q \prt_0 \Phi^{(1)} = -\frac{Q}{4}(N-1) A
\eeq
\beq{sgm9}
R^{(1)} = -4 Q \prt_0 \Phi^{(1)},
\eeq
where $\l R^{(1)}$ denotes the contracted Ricci tensor corresponding
to $G_{\m\n}$ to order $O(\l)$. $R^{(1)}$ is simply given by the
curvature dictated by the $O(N)$ model:
$$R^{(1)}=-(N-1)(N-2),$$
and \rf{sgm9} and \rf{sgm8} imply:
\beq{sgm10}
\Phi^{(1)} = \frac{1}{4Q}(N-1)(N-2)\rho,~~~~~~~~~A=\frac{2}{Q}(N-2).
\eeq

If we return to \rf{sgm1} we see that a change in the physical cut-off
$a \to a(1+dl)$, or equivalently a shift  $\rho \to \rho +2dl/\g$
(where $\g$ is given by \rf{j7a})
can be absorbed in a coupling constant renormalization
\beq{sgm11}
\l \to \l Z(\l),~~~~~~~Z(\l) = e^{\frac{4(N-2)\l dl}{\g Q}} .
\eeq
This leads to the $\b$-function:
\beq{sgm12}
\b_G(\l) = -\frac{d\l}{dl} = - \frac{4(N-2)\l^2}{\g Q} =\frac{2}{\g Q} \b(\l).
\eeq
In addition the expression \rf{sgm10} for $\Phi^{(1)}$ leads to a
renormalization of the charge $Q$:
\beq{sgm13}
\tilde{Q} =   Q + \frac{\l}{2Q} (N-1)(N-2) = Q - \frac{\l}{2Q}R^{(1)}.
\eeq
Again this is analogous to the result found in the Sine-Gordon model.

\section{Conclusion and Discussion}

We have provided some general arguments, working in conformal gauge,
in favour of a universal correction to the flat space $\b$-function
of marginal perturbations when the  theory is coupled to 2d
quantum gravity. Our conclusions agree with the recent observations
made in \cite{schmid,5}.
For three  simple models we have calculated in detail the
lowest order effective actions for 2d quantum gravity
coupled to  matter fields which in flat space have a non-trivial
$\b$-function. When the $\b$-function was derived from the effective
action we found agreement with the universal correction mentioned above.
In addition we observed that the value of $Q$ should be renormalized.
This is in agreement with general features of the renormalization group
flow from an ultra-violet to an infra-red fixed point, and in agreement
with the observation that the ``gravitational dressing'' of the
renormalization group flow seems not to  be able change
an ultra-violet stable fixed point into an infra-red stable fixed point.
While the ``dressing of the $\b$-function'' can be viewed as the first
influence of gravity on marginal operators which perturb the matter
part away from a fixed point, the change in $Q$ can be viewed as the
first back-reaction of the marginal operators on gravity itself when
we move away from the fixed point.

In the extended Sine-Gordon model  it was from a formal point
of view possible to have $c >1$ ($N >0$) if we were willing to give
up unitarity of the world surface non-linear sigma model. However, this
seems a rather dramatic modification of the theory and in fact we
found no consistent solutions to the classical equations which at the
same time allowed the interpretation as a perturbation caused by
marginal operators.

It would be interesting to calculate the next order
correction of the $\b$-function caused by the ``gravitational dressing''.
However, the arguments which used the operator product expansion can not be
generalized in a straight forward manner to the $O(\l^3)$ corrections.
Alternatively one could try first to calculate the effective action
from the classical equations of motion, but a consistent calculation
would need higher order terms of the tachyonic potential $v(T)$ as
well as higher powers of $R$. However, these terms are not universal terms
in the target space action, but depend on the renormalization scheme
used for the non-linear sigma model,
and the status of the results derived from
such a calculation is  not a priori clear to us. It seems to us a
major challenge to  clarify these issues and eventually
apply the methods to a more complicated theory like dilaton gravity.

Finally the universal gravity correction $2/\g Q$ to the $\b$-function
calls upon a geometrical interpretation directly in terms of the
fractal structure of space-time in quantum gravity. We have not yet found
this interpretation, but it is a most fascinating topic for further study.

\vspace{12pt}

\noindent {\bf Acknowledgement:} One of the author (KG) thanks
the members of high energy group in NBI for useful discussions and kindness.
We also thanks N. Sakai and A. Tseytlin for useful comments and
constructive criticism.

\vspace{24pt}





\end{document}